\documentstyle[12pt]{article}
\begin{document}
\sloppy
\begin{titlepage}
\begin{center}
{\Large\bf  BPS States in F-Theory\\}
\vspace*{15mm}
{\bf Yu. Malyuta and T. Obikhod\\}
\vspace*{10mm}
{\it Institute for Nuclear Research, National Academy of
Sciences of Ukraine\\
252022 Kiev, Ukraine\\}
{\bf E-mail: aks@d310.icyb.kiev.ua\\}
\vspace*{35mm}
{\bf Abstract\\}
\end{center}
\vspace*{1mm}
\hspace*{7mm}The spectra of BPS states in
F-theory on elliptic fibred fourfolds
are investigated.
\end{titlepage}
\section {\bf Introduction}
Klemm, Mayr and Vafa \cite{1.} have shown that F-theory
compactification on the Calabi-Yau threefold
$  X_{24}(1,1,2,8,12)  $ leads to the spectrum of
BPS states recorded in Table 1.
\begin{center}
\begin{tabular}{|l|cccrrr|}  \hline
\scriptsize $ m $&$ n_{0,0,m} $&$ n_{1,0,m} $&$ n_{2,0,m} $
&$ n_{3,0,m} $&$ n_{4,0,m} $&$ n_{5,0,m} $  \\   \hline
1   &     1       &     252     &    5130
&    54760    &   419895    &  2587788      \\
2   &             &             &\hspace*{-3mm}$ -9252 $
& $  -673760 $  & $ -20534040 $ &$ -389320128 $    \\
3   &             &             &
&    848628   &  115243155  &  6499779552   \\ \hline
\end{tabular} \\
\vspace*{3.5mm}
Table 1. Instanton numbers (degeneracies of BPS states) for
$  X_{24}(1,1,2,8,12)  $. \\
\end{center}
\vspace*{2mm}
\hspace*{7mm}Other F-theory compactifications on the special
fourfolds have been studied by Donagi, Grassi, Witten \cite{2.}
and Mayr \cite{3.} in the context of non-perturbative
superpotentials.
\hspace*{7mm}The purpose of this work is to investigate the
spectra of BPS states in F-theory on elliptic fibred fourfolds.
We consider some models which have non-perturbative
superpotentials.
\section {\bf Donagi-Grassi-Witten model}
The fourfold in this case is the elliptic fibration
over the base $ {\bf P}^{1} \times S $, where $ S $
denotes the del Pezzo surface \cite{2.}.
The model exhibits the geometrical phase which
corresponds to the regular triangulation of the
dual polihedron \cite{4.}. This phase has the following
Mori generators\\
\vspace*{-2mm}\\
\hspace*{40mm}$  l^{(1)} = (-3,0;1,1,1,0,0,0,0,0,0,0),  $
\vspace*{3mm}\\
\hspace*{40mm}$  l^{(2)} = (-1,-1;0,0,0,1,1,0,0,0,0,0), $
\vspace*{3mm}\\
\hspace*{40mm}$  l^{(3)} = (0,-2;0,0,0,0,0,1,1,0,0,0),  $
\vspace*{3mm}\\
\hspace*{40mm}$  l^{(4)} = (0,-3;0,0,0,0,0,0,0,1,1,1). $
\vspace*{3mm}\\
\hspace*{7mm}Using prescriptions of the paper \cite{5.} we derive
the principal parts of the Picard-Fuchs operators\\
\vspace*{3mm}
$ \hspace*{40mm}   L_{1} = 3\theta_{1}^{2} - \theta_{1}\theta_{2} +
\theta_{2}^{2}, $                           \\
\vspace*{3mm}
$ \hspace*{40mm}   L_{2} = \theta_{2}^{2},  $               \\
\vspace*{3mm}
$ \hspace*{40mm}   L_{3} = \theta_{3}^{2},  $               \\
\vspace*{3mm}
$ \hspace*{40mm}   L_{4} = \theta_{2}^{2} + 4\theta_{2}\theta_{3} +
4\theta_{3}^{2} - 3\theta_{2}\theta_{4} -
6\theta_{3}\theta_{4} + 9\theta_{4}^{2},  $ \\
\vspace*{3mm}
where  $   \theta_{i} = z_{i}\frac{d}{dz_{i}},
\  z_{i}  $ are the algebraic coordinates on the complex structure
moduli space.\\
\hspace*{7mm}Application of the program INSTANTON  \cite{6.} gives
the spectrum of BPS states recorded in Table 2.
\begin{center}
\begin{tabular}{|llccrr|}    \hline
$ n_{2,1,3,0,0} $&$ n_{2,1,3,0,1} $&$ n_{2,1,3,0,2} $&
$ n_{2,1,3,0,3} $  &$ n_{2,1,3,0,4} $&$ n_{2,1,3,0,5} $ \\  \hline
\hspace*{5mm}1   &\hspace*{3mm}252 &      5130       &
       54760       &     419895      &   2587788        \\
                 &                 &                 &
    (27 sequences) &                 &
                  \\  \hline  \hline
                   &$ n_{2,1,3,0,1} $&$ n_{2,2,3,0,1} $&
$ n_{2,3,3,0,1} $  &$ n_{2,4,3,0,1} $&$ n_{2,5,3,0,1} $ \\  \hline
                   &$ 1\cdot252^{\ast} $&$ 2\cdot5130^{\ast} $&
$ 3\cdot54760^{\ast} $&$ 4\cdot419895^{\ast} $&
$ 5\cdot2587788^{\ast} $  \\
                 &                 &                 &
    (27 sequences) &                 &
                  \\  \hline   \hline
$ n_{4,0,3,0,1} $&$ n_{4,1,3,0,1} $&$ n_{4,2,3,0,1} $&
$ n_{4,3,3,0,1} $  &$ n_{4,4,3,0,1} $&$ n_{4,5,3,0,1} $ \\  \hline
\hspace*{5mm}1   &\hspace*{3mm}252 &      5130       &
       54760       &     419895      &   2587788        \\
                 &                 &                 &
   (513 sequences) &                 &
                  \\  \hline   \hline
$ n_{5,0,3,0,1} $&$ n_{5,1,3,0,1} $&$ n_{5,2,3,0,1} $&
$ n_{5,3,3,0,1} $  &$ n_{5,4,3,0,1} $&$ n_{5,5,3,0,1} $ \\  \hline
\hspace*{5mm}1   &\hspace*{3mm}252 &      5130       &
       54760       &     419895      &    2587788       \\
                 &                 &                 &
  (702 sequences)  &                 &
                  \\  \hline   \hline
$ n_{6,0,3,0,1} $&$ n_{6,1,3,0,1} $&$ n_{6,2,3,0,1} $&
$ n_{6,3,3,0,1} $  &$ n_{6,4,3,0,1} $&$ n_{6,5,3,0,1} $ \\  \hline
\hspace*{5mm}1   &\hspace*{3mm}252 &      5130       &
       54760       &     419895      &    2587788       \\
                 &                 &                 &
  (189 sequences)  &                 &                  \\  \hline
\end{tabular}\\
\vspace*{3.5mm}
Table 2. Instanton numbers (degeneracies of BPS states) for the
Donagi-Grassi-Witten\\
\hspace*{-113mm}model.
\end{center}
\hspace*{7mm}We found the tower of infinite sequences of BPS
degeneracies. The sequences 1, 252, 5130, {\ldots} are well-known.
The $ E_{8} $ partition function is the counting functional
of these sequences \cite{7.}. But the sequences
$ 1\cdot252 $,\hspace*{1mm}$ 2\cdot5130 $,
$ 3\cdot54760 $, {\ldots} (marked by the $ {\ast} $ in Table 2)
are unfamiliar, and  physically interesting, as they
signal the appearance of additional states.
\section {\bf Elliptic fibred fourfold over the $ {\bf P}^{1} $-
bundle on $ ({\bf P}^{1})^{2} $}
The fourfold in this case is the elliptic fibration over the
$ {\bf P}^{1} $-bundle $ {\bf P}({\cal O}_{B}\otimes{\cal O}_{B} $
$ (f_{1}+f_{2})) $ on $ B = ({\bf P}^{1})^{2} $, where
$ f_{1} $ and $ f_{2} $ are fibres of the two projections from
$ B $ to $ {\bf P}^{1} $ \cite{3.}\cite{7.}. This model exhibits
the geometrical phase which has the Mori generators\\
\vspace*{-2mm}\\
\hspace*{40mm}$   l^{(1)} = (0;1,0,0,0,0,-1,0,-1,1), $
\vspace*{3mm}\\
\hspace*{40mm}$   l^{(2)} = (0;0,0,1,0,0,1,0,-2,0), $
\vspace*{3mm}\\
\hspace*{40mm}$   l^{(3)} = (-6;0,0,0,2,3,0,0,1,0),  $
\vspace*{3mm}\\
\hspace*{40mm}$   l^{(4)} = (0;0,1,0,0,0,-1,1,-1,0). $
\vspace*{3mm}\\
\hspace*{7mm}The principal parts of the Picard-Fuchs
operators are\\
\vspace*{-2mm}\\
\hspace*{40mm} $  L_{1} = \theta_{1}^{2} , $
\vspace*{3mm}\\
\hspace*{40mm} $  L_{2} = \theta_{2}(\theta_{1}-\theta_{2}
+\theta_{4}),  $
\vspace*{3mm}\\
\hspace*{40mm} $  L_{3} = \theta_{3}(\theta_{1}+2\theta_{2}
-\theta_{3}+\theta_{4}),  $
\vspace*{3mm}\\
\hspace*{40mm} $  L_{4} = \theta_{4}^{2}.   $
\vspace*{3mm}\\
\hspace*{7mm}With the program INSTANTON we obtain the spectrum of
BPS states recorded in Table 3.\\
\begin{center}
\begin{tabular}{|llccrr|}    \hline
$ n_{3,0,0,0,1} $&$ n_{3,0,0,1,1} $&$ n_{3,0,0,2,1} $&
$ n_{3,0,0,3,1} $  &$ n_{3,0,0,4,1} $&$ n_{3,0,0,5,1} $ \\  \hline
\hspace*{5mm}1   &\hspace*{3mm}252 &      5130       &
       54760       &     419895      &   2587788        \\
                 &                 &                 &
(1 sequence)     &                 &
                  \\  \hline  \hline
$ n_{3,1,0,0,0} $&$ n_{3,1,0,1,0} $&$ n_{3,1,0,2,0} $&
$ n_{3,1,0,3,0} $  &$ n_{3,1,0,4,0} $&$ n_{3,1,0,5,0} $ \\  \hline
\hspace*{5mm}1   &\hspace*{3mm}252 &      5130       &
       54760       &     419895      &   2587788        \\
                   &                 &                 &
  (1 sequence)     &                 &
                  \\  \hline  \hline
                   &                 &$ n_{3,0,0,2,2} $&
$ n_{3,0,0,3,2} $  &$ n_{3,0,0,4,2} $&$ n_{3,0,0,5,2} $ \\  \hline
                   &                 &    $  -9252 $     &
 $    -673760 $    & $   -20534040 $ & $  -389320128 $    \\
                   &                 &                 &
     (2 sequences) &                 &
                  \\  \hline  \hline
                   &                 &$ n_{3,2,0,2,0} $&
$ n_{3,2,0,3,0} $  &$ n_{3,2,0,4,0} $&$ n_{3,2,0,5,0} $ \\  \hline
                   &                 &    $  -9252 $     &
  $   -673760 $    & $   -20534040 $ & $  -389320128 $  \\
                   &                 &                 &
     (2 sequences) &                 &
                  \\  \hline  \hline
                   &                 &                 &
$ n_{3,0,0,3,3} $  &$ n_{3,0,0,4,3} $&$ n_{3,0,0,5,3} $ \\  \hline
                   &                 &                 &
      848628       &    115243155    &    6499779552     \\
                   &                 &                 &
     (3 sequences) &                 &
                  \\  \hline  \hline
                   &                 &                 &
$ n_{3,3,0,3,0} $  &$ n_{3,3,0,4,0} $&$ n_{3,3,0,5,0} $ \\  \hline
                   &                 &                 &
      848628       &    115243155    &    6499779552     \\
                   &                 &                 &
     (3 sequences) &                 &
                  \\  \hline  \hline
                   &$ n_{5,1,0,1,0} $&$ n_{5,1,0,2,0} $&
$ n_{5,1,0,3,0} $  &$ n_{5,1,0,4,0} $&$ n_{5,1,0,5,0} $ \\  \hline
                   &$ -1\cdot252^{\ast} $&$ -2\cdot5130^{\ast} $&
$ -3\cdot54760^{\ast} $&$ -4\cdot419895^{\ast} $&
$ -5\cdot2587788^{\ast} $               \\
                   &                 &                 &
     (2 sequences) &                 &
                  \\  \hline  \hline
                   &                 &$ n_{5,2,0,2,0} $&
$ n_{5,2,0,3,0} $  &$ n_{5,2,0,4,0} $&$ n_{5,2,0,5,0} $ \\  \hline
                   &                 &$ 2\cdot9252^{\ast} $&
$ 3\cdot673760^{\ast} $&$ 4\cdot20534040^{\ast} $&$
5\cdot389320128^{\ast} $            \\
                   &                 &                 &
     (2 sequences) &                 &
                  \\  \hline  \hline
                   &                 &                 &
$ n_{5,3,0,3,0} $  &$ n_{5,3,0,4,0} $&$ n_{5,3,0,5,0} $ \\  \hline
                   &                 &                 &
$ -3\cdot848628^{\ast} $&$ -4\cdot115243155^{\ast} $&
$ -5\cdot6499779552^{\ast} $        \\
                   &                 &                 &
     (2 sequences) &                 &
                  \\       \hline
\end{tabular}\\
\vspace*{3.5mm}
Table 3. Instanton numbers (degeneracies of BPS states) for the
elliptic fibred fourfold \\
\hspace*{-70mm}over the $ {\bf P}^{1} $-
bundle on $ ({\bf P}^{1})^{2} $.
\end{center}
\newpage
\hspace*{1mm}The sequences\\
\hspace*{51.8mm} 1, 252, 5130, {\ldots},    \\
\hspace*{50mm}$ -9252, -673760, -20534040,$ {\ldots},    \\
\hspace*{53mm}848628, 115243155, 6499779552, {\ldots}    \\
are well-known. But the sequences \\
\hspace*{50mm}$ -1\cdot252 $,\hspace*{1mm}$ -2\cdot5130 $,
$ -3\cdot54760 $, {\ldots},    \\
\hspace*{53mm}$ 2\cdot9252 $,\hspace*{1mm}$ 3\cdot673760 $,
$ 4\cdot20534040 $, {\ldots},    \\
\hspace*{50mm}$ -3\cdot848628 $,\hspace*{1mm}$ -4\cdot115243155 $,
$ -5\cdot6499779552 $, {\ldots}    \\
(marked by the $ {\ast} $ in Table 3) are unfamiliar, and
physically interesting, as they signal the appearence of
additional states.
\section
{\bf Elliptic fibred fourfold over $ {\bf P}^{1}\times F_{1} $}
The fourfold in this case is the elliptic fibration over
$ {\bf P}^{1}\times F_{1} $, where  $ F_{1} $ is the Hirzebruch
surface \cite{3.}\cite{7.}. This model exhibits the
geometrical phase which has the Mori generators \\
\vspace*{-2mm}\\
\hspace*{40.5mm} $ l^{(1)} = (-6;0,0,0,0,0,0,2,3,1),  $
\vspace*{3mm}\\
\hspace*{41.5mm}$  l^{(2)} = (0;0,0,0,0,1,1,0,0,-2), $
\vspace*{3mm}\\
\hspace*{40mm} $ l^{(3)} = (0;0,1,0,1,0,-1,0,0,-1),  $
\vspace*{3mm}\\
\hspace*{40mm} $ l^{(4)} = (0;1,-2,1,0,0,0,0,0,0). $
\vspace*{3mm}\\
\hspace*{7mm}The principal parts of the Picard-Fuchs operators are\\
\vspace*{-2mm}\\
\hspace*{41.5mm}$   L_{1} = \theta_{1}(\theta_{1}-2\theta_{2}
-\theta_{3}) , $
\vspace*{3mm}\\
\hspace*{41.5mm}$   L_{2} = \theta_{2}(\theta_{2}-\theta_{3}), $
\vspace*{3mm}\\
\hspace*{40mm} $  L_{3} = \theta_{3}(\theta_{3}-2\theta_{4}), $
\vspace*{3mm}\\
\hspace*{40mm} $  L_{4} = \theta_{4}^{2} .  $
\vspace*{3mm}\\
\hspace*{7mm}With the program INSTANTON we obtain the spectrum
of BPS states recorded in Table 4.
\begin{center}
\begin{tabular}{|llccrr|}    \hline
                   &$ n_{3,1,0,1,0} $&$ n_{3,2,0,1,0} $&
$ n_{3,3,0,1,0} $  &$ n_{3,4,0,1,0} $&$ n_{3,5,0,1,0} $ \\  \hline
                   &$ 1\cdot252^{\ast} $&$ 2\cdot5130^{\ast} $&
$ 3\cdot54760^{\ast} $&$ 4\cdot419895^{\ast} $&
$ 5\cdot2587788^{\ast} $               \\
                   &                 &                 &
      (1 sequence) &                 &
                  \\  \hline  \hline
                   &                 &$ n_{3,2,0,2,0} $&
$ n_{3,3,0,2,0} $  &$ n_{3,4,0,2,0} $&$ n_{3,5,0,2,0} $ \\  \hline
                   &                 &$ -2\cdot9252^{\ast} $&
$ -3\cdot673760^{\ast} $&$ -4\cdot20534040^{\ast} $&$
-5\cdot389320128^{\ast} $            \\
                   &                 &                 &
      (1 sequence) &                 &
                  \\  \hline  \hline
                   &                 &                 &
$ n_{3,3,0,3,0} $  &$ n_{3,4,0,3,0} $&$ n_{3,5,0,3,0} $ \\  \hline
                   &                 &                 &
$ 3\cdot848628^{\ast} $&$ 4\cdot115243155^{\ast} $&
$ 5\cdot6499779552^{\ast} $        \\
                   &                 &                 &
      (1 sequence) &                 &
                  \\       \hline
$ n_{6,0,0,1,0} $&$ n_{6,1,0,1,0} $&$ n_{6,2,0,1,0} $&
$ n_{6,3,0,1,0} $  &$ n_{6,4,0,1,0} $&$ n_{6,5,0,1,0} $ \\  \hline
\hspace*{5mm}1   &\hspace*{3mm}252 &      5130       &
       54760       &     419895      &   2587788        \\
                   &                 &                &
 (5 sequences)     &                 &
                  \\  \hline  \hline
$ n_{6,0,0,1,1} $&$ n_{6,1,0,1,1} $&$ n_{6,2,0,1,1} $&
$ n_{6,3,0,1,1} $  &$ n_{6,4,0,1,1} $&$ n_{6,5,0,1,1} $ \\  \hline
\hspace*{5mm}1   &\hspace*{3mm}252 &      5130       &
       54760       &     419895      &   2587788        \\
                   &                 &                &
 (5 sequences)     &                 &
                  \\  \hline  \hline
                   &                 &$ n_{6,2,0,2,0} $&
$ n_{6,3,0,2,0} $  &$ n_{6,4,0,2,0} $&$ n_{6,5,0,2,0} $ \\  \hline
                   &                 &   $   -9252 $     &
  $   -673760 $    & $   -20534040 $ & $  -389320128 $     \\
                   &                 &                 &
    (10 sequences) &                 &
                  \\  \hline  \hline
                   &                 &$ n_{6,2,0,2,2} $&
$ n_{6,3,0,2,2} $  &$ n_{6,4,0,2,2} $&$ n_{6,5,0,2,2} $ \\  \hline
                   &                 &    $  -9252 $     &
  $   -673760 $    & $   -20534040 $ & $  -389320128  $  \\
                   &                 &                 &
    (10 sequences) &                 &
                  \\  \hline  \hline
                   &                 &                 &
$ n_{6,3,0,3,0} $  &$ n_{6,4,0,3,0} $&$ n_{6,5,0,3,0} $ \\  \hline
                   &                 &                 &
      848628       &    115243155    &    6499779552     \\
                   &                 &                 &
    (15 sequences) &                 &
                  \\  \hline  \hline
                   &                 &                 &
$ n_{6,3,0,3,3} $  &$ n_{6,4,0,3,3} $&$ n_{6,5,0,3,3} $ \\  \hline
                   &                 &                 &
      848628       &    115243155    &    6499779552     \\
                   &                 &                 &
    (15 sequences) &                 &
                  \\   \hline
\end{tabular}\\
\vspace*{3.5mm}
Table 4. Instanton numbers (degeneracies of BPS states) for the
elliptic fibred fourfold\\
\hspace*{-100mm}over $ {\bf P}^{1}\times F_{1} $.
\end{center}
\hspace*{7mm}The sequences\\
\hspace*{51.8mm} 1, 252, 5130, {\ldots},    \\
\hspace*{50mm}$ -9252, -673760, -20534040,$ {\ldots},    \\
\hspace*{53mm}848628, 115243155, 6499779552, {\ldots}    \\
are well-known. But the sequences \\
\hspace*{53.2mm}$ 1\cdot252 $,\hspace*{1mm}$ 2\cdot5130 $,
$ 3\cdot54760 $, {\ldots},    \\
\hspace*{50mm}$ -2\cdot9252 $,\hspace*{1mm}$ -3\cdot673760 $,
$ -4\cdot20534040 $, {\ldots},    \\
\hspace*{53.8mm}$ 3\cdot848628 $,\hspace*{1mm}$ 4\cdot115243155 $,
$ 5\cdot6499779552 $, {\ldots}    \\
(marked by the $ {\ast} $ in Table 4) are unfamiliar, and
physically interesting, as they signal the appearence of
additional states.

\end{document}